\documentclass[fleqn]{article}
\usepackage{amsmath,amssymb}
\usepackage[dvips]{graphicx}

\usepackage{latexsym}

\begin{document}

\pagestyle{plain} 
\setcounter{page}{1}
\setlength{\textheight}{700pt}
\setlength{\topmargin}{-40pt}
\setlength{\headheight}{0pt}
\setlength{\marginparwidth}{-10pt}
\setlength{\textwidth}{20cm}

\title{Does Parrondo Paradox occur in Scale Free Networks?\\--A simple Consideration--}
\author{Norihito Toyota   \and Hokkaido Information University, Ebetsu, Nisinopporo 59-2, Japan \and email :toyota@do-johodai.ac.jp }
\date{}
\maketitle

\begin{abstract}
Parrondo's paradox occurs in sequences of games in which a winning expectation may be obtained by playing the games in a random order, even though each game in the sequence may be lost when played individually. 
Several variations of Parrondo's games apparently with paradoxical property have been introduced;
 history dependence, one dimensional line,  two dimensional lattice and so on. 
In this article, we examine whether Parrondo's paradox occurs or not in scale free networks. 
This  is interesting as an empirical study, since scale free networks are ubiquitous in our real world. 
First some simulation results are given and after that theoretical studies are made. 
As a result, we mostly confirm that Parrondo's paradox can not occur in the naive case, where the game has the same number of parameters 
as the original Parrondo's game.   

 \end{abstract}
\begin{flushleft}
\textbf{keywords:}
 Parrondo's paradox, Parrondo's paradox，Scale  free network, Game theory
\end{flushleft}

\section{Introduction}\label{intro}
\hspace{5mm} 
Parrondo's games were first devised by Parrondo \cite{Parr1} who presented them in unpublished form and represent a sort of coin flipping game.  
Parrondo's paradox is based on the combination of two negatively biased or losing game. However, when the two games are randomly combined, they give rise to a positively biased game or a winning game pointed out by \cite{Harm1} and \cite{Harm2}. 
This paradox is a translation of the physical mode of the flashing Brownian ratchet to game theoretical one\cite{Amen}.       
It was also pointed out that Fokker Planck equation connect  Parrondo's games with the flashing Brownian ratchet \cite{Amen}. 
The original Parrondo's game consists of two losing games A and B where each are played by only one player.  
In the game A, only one biased coin is used, while in the game B two biased coins are used with the player's current capital determining the state dependent rule. When a player plays individually each game, he/she  loses his/her capital on average. 
However, when the player plays two games in any combination, he/she always wins on average.    

Several variations of Parrondo's games apparently with paradoxical property have been introduced. 
First capital dependence in the game B was replaced by recent history of wins and loses \cite{Parr2}. 
It has the same paradoxical property as the original one. 
For it, an analytical study has been made \cite{Rasm1}. 
Secondly the capital dependence in the game B was replaced by spatial neighbor dependence in one dimensional line \cite{Toral1}, \cite{Miha1}, \cite{Miha2},  and von Neumann  neighbor dependence in two dimensional lattice \cite{Miha}.  
Some analogous variations have been given by  \cite{Toral2},\cite{Davi}. 
A fine review of Parrondo's paradox and references are given in \cite{Harm3}. 

In this article, we examine whether Parrondo's paradox occurs or not in scale free networks\cite{Albe1},\cite{Albe2}, instead of two dimensional lattice.
This  is interesting as an empirical study, since scale free networks are a common  occurrence in our real world\cite{newBook}. 
First some simulation results are given and after that theoretical studies are made. 
As a result, we mostly confirm that Parrondo's paradox can not occur in the naive case, where each player on a network plays a game L when there are $R$ or more winners in the neighborhood connected to the player and plays a game W otherwise in the game B.

\section{Review of Parrondo's Paradox}
\subsection{Original Game}
\hspace{5mm} 
We briefly review the original Parrondo's game in this section. 
Parrondo's paradox occurs in sequences of games in which a winning expectation may be obtained by playing the games in a random order, even though each game in the sequence may be lost when played individually.  
The original version of Parrondo's game consists of  the following two games and the initial capital of a player is $C(0)=0$; \\
\begin{itemize}
\item Game A: the probability  of winning is $P_A$ in this game. Usually $P_A<0.5$ is taken for losing game. It is played by using a biased coin. 
\item Game B: If the capital $C(t)$ of the player at $t$ is a multiple of 3, the probability of winning is $P_B^{(1)}$, otherwise, the probability of winning is $P_B^{(2)}$
\item Game A+B: Two games are mixed. The game A is played with probability $P$ and game B is played with $1-P$.
\end{itemize}
In all there are 4 parameters,  $P$, $P_A$, $P_B^{(1)}$ and $P_B^{(2)}$,  controllable by  a planner of the game  in Parrondo's game.  
When we win a game A or B,  we get one unit of capital and when we lose the game, we lose one unit of capital.

Two time series of $C(t)$ of the game A and the game B are given by Fig.1, respectively. 
They show  that  the capital $C(t)$ decreases with time $t$ in both games. 
Fig 2, which displays the time series of the gama A+B, however, shows that   the capital $C(t)$ increases with time $t$, 
where $C(0)=100$ is taken. 
This is an example of Parrondo's paradox. 

 \begin{figure}[t]
\begin{center}
\includegraphics[scale=0.7,clip]{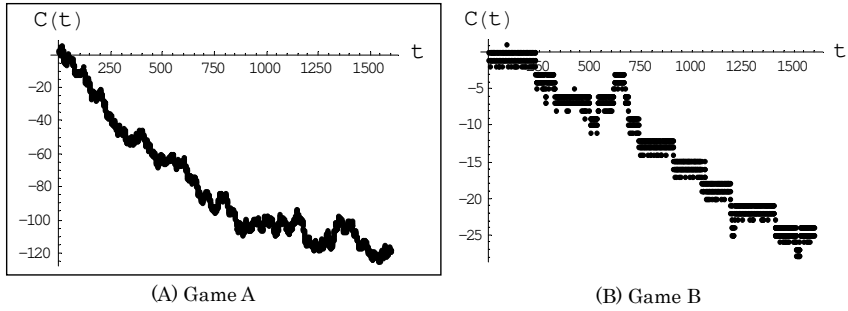} 
\end{center}
\caption{$C(t)$ in Game A and Game B for $P_A=0.48$,  $P_B^{(2)}=0.01$ and $P_B^{(1)}=0.85$ at 1600 steps.  }
\vspace{4mm}
\end{figure}

 \begin{figure}[t]
\begin{center}
\includegraphics[scale=0.6,clip]{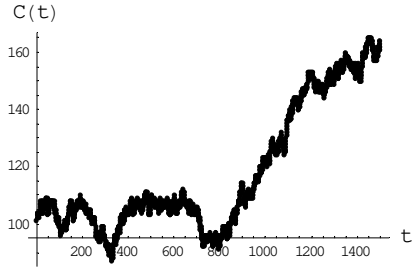} 
\end{center}
\caption{$C(t)$ in Game A+B for $P=0.5$, $P_A=0.48$,  $P_B^{(2)}=0.01$ and $P_B^{(1)}=0.85$ at 1600 steps.}
\end{figure} 
 
\subsection{Theoretical Studies}
\hspace{5mm} 
We give theoretical considerations in Parrondo's game in this subsection and derive a condition 
under which the paradox occurs.  
Then we follow the discussions developed in \cite{Amen}.  

First of all, we start with the analysis of the game A. 
Let $f_j$  be the probability that the capital reaches zero in a finite number of plays, 
supposing that we  initially have a given capital of $j$ units. 
\begin{itemize}
\item $f_j \geq1$ for all $j \geq 0$,  then the game A is either fair or losing one.  
\item $f_j<1$ for all $j > 0$,  then the game A can be a winning one  because there is a possibility that our capital can grow indefinitely. 
\end{itemize}

We calculate $f_j$ from the following recursion relation;
\begin{equation}
f_j = P_A \cdot f_{j+1} + (1-P_A)\cdot f_{j-1}, \;\;\; j\geq 1 
\end{equation} 
with the initial condition
\begin{equation}
f_0 = 1.  
\end{equation} 
The solution of eq. (1) for the initial condition (2) is 
\begin{equation}
f_j = \alpha \Bigr[ \Bigr( \frac{1-P_A}{P_A} \Bigl) ^j    -1\Bigl] +1,  
\end{equation} 
where $\alpha$ is a constant.  
For the minimal non-negative solution, we obtain 

\begin{equation}
f_j = \min \Bigr[ \Bigr( \frac{1-P_A}{P_A} \Bigl) ^j,  1 \Bigl].   
\end{equation} 
Then we find that 
\begin{eqnarray}
\frac{1-P_A}{P_A} < 1  &so& P_A>\frac{1}{2}\;\;\;\;\;\mbox{ winning game},\\
\frac{1-P_A}{P_A} = 1    &so& P_A=\frac{1}{2} \;\;\;\;\;\mbox{ fair game},\\
\frac{1-P_A}{P_A} > 1  &so &  P_A<\frac{1}{2}    \;\;\;\;\;\mbox{ losing game}.
\end{eqnarray} 

Second we consider the game B. 
Let $g_i$ be the probability that the capital will reach zero in a finite number of plays,  
supposing that we  initially have a given capital of $j$ units. 
As in the game A, 
\begin{itemize}
\item $g_j=1$ for all $j \geq 0$,  then the game B is either fair or losing one.  
\item $g_j<1$ for all $j > 0$,  then the game B can be a winning one  because there is a possibility that our capital can grow indefinitely. 
\end{itemize}
On this occasion, we have to solve the following recursion relations;
\begin{eqnarray}
g_{3j} &=&P_B^{(1)} \cdot g_{3j+1}+ (1-P_B^{(1)} )\cdot g_{3j-1}  \;\;\;\;\;j\geq 1, \nonumber \\
g_{3j+1} &=&P_B^{(2)} \cdot g_{3j+2}+ (1-P_B^{(2)} )\cdot g_{3j}  \;\;\;\;\;j\geq 0 ,\nonumber \\
g_{3j+2} &=&P_B^{(2)} \cdot g_{3j+3}+ (1-P_B^{(2)} )\cdot g_{3j+1}  \;\;\;\;\;j \geq 0. 
\end{eqnarray} 
Eliminating $g_{3j+1} $ and $g_{3j-1} $ from eq.(8), 
we obtain
\begin{equation}
\bigr[ P_A(P_B^{(2)})^2 +(1-P_B^{(1)})(1-P_B^{(2)})^2 \bigl]\cdot g_{3j}  =P_A(P_B^{(2)})^2  \cdot g_{3j+3} + (1-P_B^{(1)})(1-P_B^{(2)})^2  \cdot g_{3j-3}. 
\end{equation} 
Considering the same initial condition $g_0=1$ as one in the game A, the general solution of eq.(9) is given by 
\begin{equation}
 g_{3j}  = \beta \Bigr[\Bigr( \frac{ (1-P_B^{(1)})(1-P_B^{(2)})^2}{ P_A(P_B^{(2)})^2 } \Bigl)^j -1 \Big]+1,  
\end{equation} 
where $\beta$ is a constant. As in the game A, for the minimal non-negative solution we get 
\begin{equation}
 g_{3j}  = \min \ \Bigr[\Bigr( \frac{ (1-P_B^{(1)})(1-P_B^{(2)})^2}{ P_A(P_B^{(2)})^2 } \Bigl)^j, 1 \Big].   
\end{equation} 
The same solution as eq. (11) is also obtained by solving eq.(8) with respect to $ g_{3j+1} $ and $ g_{3j+2} $.  
As the game A, we obtain 
\begin{eqnarray}
\frac{ (1-P_B^{(1)})(1-P_B^{(2)})^2}{ P_A(P_B^{(2)})^2 } < 1  &\;\;\;\;\;\mbox{ winning game},\\
\frac{ (1-P_B^{(1)})(1-P_B^{(2)})^2}{ P_A(P_B^{(2)})^2 } = 1    &\;\;\;\;\mbox{ fair game},\\
\frac{ (1-P_B^{(1)})(1-P_B^{(2)})^2}{ P_A(P_B^{(2)})^2 }  > 1  &\;\;\;\;\mbox{ losing game}.
\end{eqnarray} 

Now we will turn the game A+B where the game A is chosen randomly with the probability $P$ or the game B  with the probability $1-P$. 
We classify the game into two cases  as in the game B.   
   \begin{itemize}
\item The case that the capital is a multiple of three.\\
        The winning probability is 
\begin{equation}
 P_{A+B}^{(1)} \equiv  PP_A + (1-P)P_B^{(1)}.   
\end{equation}  
\item The case that the capital is not a multiple of three\\
        The winning probability is 
\begin{equation}
 P_{A+B}^{(2)} \equiv  PP_A + (1-P)P_B^{(2)}.   
\end{equation}  
\end{itemize}

So the game A+B is included in the game B as a special case by the following replacement;  
\begin{equation}
 P_{B}^{(1)} \Longrightarrow  P_{A+B}^{(1)} \;\;\;\;\;\;P_{B}^{(2)} \Longrightarrow  P_{A+B}^{(2)}.   
\end{equation}  
When we define the following $ D_{(A+B)}^{(3)} $,  
\begin{equation}
  D_{(A+B)}^{(3)} \equiv \frac{ (1-P_{A+B}^{(1)})(1-P_{A+B}^{(2)})^2}{ P_{A+B}^{(1)}(P_{A+B}^{(2)})^2 } 
\end{equation}
the game A+B is classified  such as the game B 
\begin{eqnarray}
 D_{(A+B)}^{(3)}  < 1  &\;\;\mbox{ winning},\\
 D_{(A+B)}^{(3)} = 1    & \;\;\mbox{ fair},\\
 D_{(A+B)}^{(3)} > 1  &\;\;\mbox{ losing }.
\end{eqnarray} 

The game A can be also considered as the game B with $P_B^{(1)}=P_B^{(2)}=P_A$. 
Considering it together with eq. (17), both the game A and the game A+B are essentially special cases of the game B.  
Thus we can represent the discriminants in the unified way by replacing the subscript B with A or A+B  in eq. (12)-(14).

When the conditions 
\begin{eqnarray}
 D_{(A)}^{(3)}  > 1, &
 D_{(B)}^{(3)} >  1, & 
 D_{(A+B)}^{(3)} < 1 
\end{eqnarray} 
are simultaneously satisfied, Parrondo's paradox occurs.

Let consider the a more general case that  when a capital satisfies  $mod(C(t),M)= 0$ for a positive integer $M$ in the game B, 
the game with a winning probability $P_B^{(1)}$ is played and the another  game with a winning probability $P_B^{(2)}$ is played 
 in the cases of  $mod(C(t),M) \neq 0$ in the game B. 
Then the  discriminant is given by 
\begin{equation}
  D_{A+B}^{(n)}= \frac{ (1-P_{A+B}^{(1)})(1-P_{A+B}^{(2)})^{M-1}}{ P_{A+B}^{(1)}(P_{A+B}^{(2)})^{M-1} }.     
\end{equation}
Then it is known that the paradox occurs under the conditions like eq. (22)\cite{Harm3}.

\subsection{Some Extensions of Parrondo's Paradox }
\hspace{5mm} 
Such paradox also occurs in some extended versions of the original Parrondo's game. 
One of them is that two games  in the game B are chosen depending on historical information as to winning-losing  
\cite{Parr2}. 
Some theoretical discussion are given for the game\cite{Rasm1}. 

The case that players lying on one dimensional circle lattice play Parrondo's game as an other case\cite{Toral1},\cite{Miha1},\cite{Miha2} is investigated. 
Then two games in the game B are chosen depending on information as to winning-losing of two adjacent players ( left and right of the target player). 
Moreover the case is extended to the one where players lie on two dimensional regular lattice with the periodic boundary condition\cite{Miha}.  
It has been verified that the paradox can also occur in these cases\cite{Toral1},\cite{Miha}. 

For the subsequent discussions, we briefly review the extended case to two dimensional lattice according to Mihailovic et al. \cite{Miha}. 
They introduce five subgames in the game B, depending on one winner, two winners, $\cdots$ four winners in von Neumann neighborhood of a target player. 
The target player plays one of games with the winning probability of $p_B^{(0)}$，$p_B^{(1)}$，$p_B^{(2)}$，$p_B^{(3)}$ and $p_B^{(4)}$ corresponding to 
 the number of winners of his/her adjacent player. 
So the game B has five parameters. 
They analyze  the average capital over all players as the time series of capitals.  
As result, they report that the paradoxical phenomena occur in wide range of parameter sets, especially it appears typically in synchronous cases where all players play at the same time.   

\section{Parrondo's Paradox on Scale Free Networks}

\subsection{Parrondo's Game on Scale Free Networks}
\hspace{5mm} 
The degree distribution of scale free networks is given by 
\begin{equation}
P(k)=ck^{-\alpha},
\end{equation}
where $\alpha>0$ an exponent in a power low. 
The normalization constant $c$ is determined by   
\begin{equation}
\sum_{k=k_{min}}^{k_{max}} cP(k)= \sum_{k=k_{min}}^{k_{max}} ck^{-\alpha} =1, 
\end{equation}
where $k_{min} $ and $k_{max} $ are the maximum degree and the minimum one in a network, and $k_{min} =0$ and $k_{max} =\infty$ in an ideal case. 
$k_{min}>0$ is imposed to reduce the divergence in a realistic case.
Using the Hurwitz zeta function
\begin{equation}
\zeta (\alpha, k_{min}) \equiv \sum_{n=0}^\infty (n+k_{min})^{-\alpha}, 
\end{equation}
the degree distribution is given by 
\begin{equation}
P(k)= \frac{k^{-\alpha}}{\zeta (\alpha, k_{min})}.
\end{equation}
In the continuous limit, the normalization constant $c$ is determined as  
\begin{equation}
\int_{k_{min}}^{k_{max}} P(k) dk =\int_{k_{min}}^{k_{max}}   ck^{-\alpha}dk =1. 
\end{equation}
Then the degree distribution is 
\begin{equation}
P(k)= \frac{(\alpha -1)}{ k_{min}^{1-\alpha} - k_{max}^{1-\alpha} }\;\; k^{-\alpha} . 
\end{equation}

In this article, we construct scale free networks by using the preferential attachment (BA model) introduced by Barabashi et al. \cite{Albe1},\cite{Albe2} 
where the scaling exponent is fixed at  $\alpha=3$.      
Begin with the complete graph with degree 4,  nodes with degree 4 to the initial graph are attached by using BA model up to the network size of $N$. 
Then we expect $ k_{min}=4$ and can estimate $ k_{max}$ from $k_{min}=4$  and eq.(29). 
When the number of nodes with large degree becomes less than $one$ in the degree distribution eq.(29), it is expected that the node has about $ k_{max}$. 
Take accounting of $\alpha=3$,  $ k_{max}$ can be estimated from 
\begin{equation}
NP(k)= \frac{2N}{ 4^{-2} - k_{max}^{-2} }\;\; k_{max}^{-3} <1.
\end{equation}
That is, $ k_{max}$ is the minimum positive integer that satisfies 
\begin{equation}
 k_{max}^{3} -16 k_{max} -32N>0. 
\end{equation}

We consider a naive extension of Parrondo's game for games on scale free networks. 
When Parrondo's game is straightforwardly extended to the game with degree $k$  according to \cite{Miha}, 
the number of parameters becomes $k+1$ and that is so large. 
Furthermore the degree differs in every node in scale free networks, and so by all accounts, the game has too many parameters 
to analyze the game theoretically. 
So we introduce a cutoff as a simple idea to analyze the game on scale free networks.   
When there are not less than $R$ winners in players adjacent to a target player, the target player plays the game W whose winning probability is $P_W$, 
 and the target player plays the game L whose winning probability is $P_L $ in other cases  in the game B. 
The number of parameters of the games is the same as the original Parrondo's game due to this simplification.  
Since Parrond like paradox may accidentally occur in the capital of individual because of probability game,  
we analyze the average capital over all players as the time series of capitals in similar manner to the game on two dimensional lattice.

 \begin{figure}[t]
\begin{center}
\includegraphics[scale=0.9,clip]{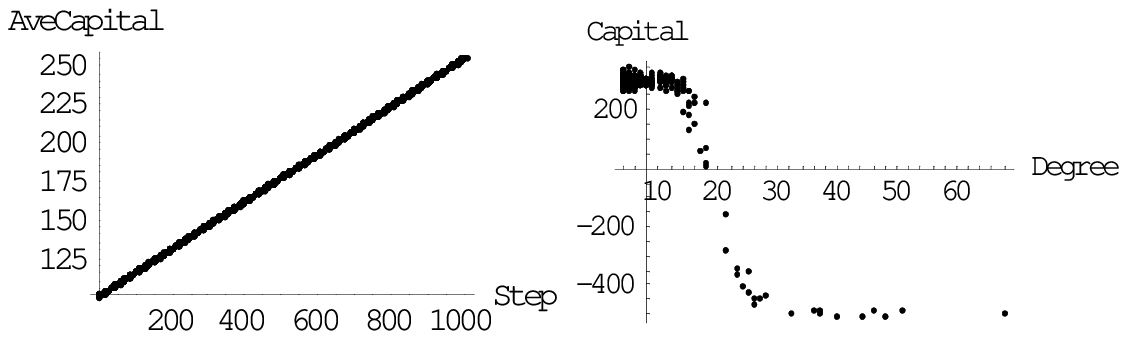} 
\end{center}
\caption{ Average capital $C(t)$ and capital distribution in every node in the game B at $R=10$ and $(P_L, P_R)=(0.6,0.2)$.  }
\vspace{4mm}
\begin{center}
\includegraphics[scale=0.9,clip]{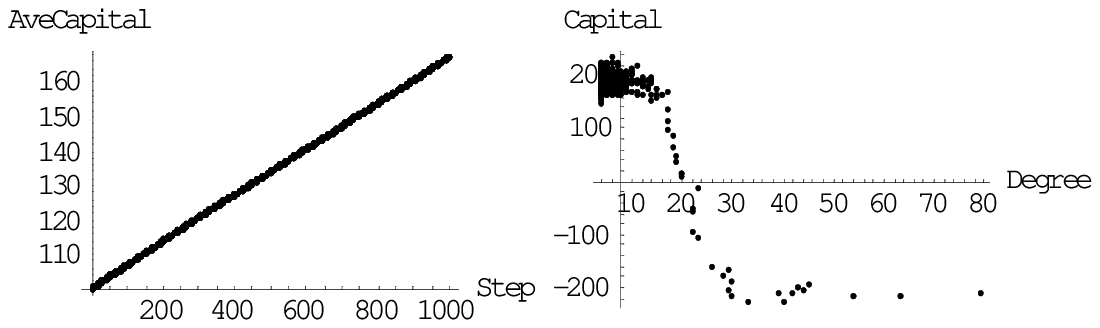} 
\end{center}
\caption{ Average capital $C(t)$ and capital distribution in every node in the game A+B at $R=10$ and$(P_L, P_R)=(0.6,0.2)$.  }
\end{figure}

 \begin{figure}[t]
\begin{center}
\includegraphics[scale=0.9,clip]{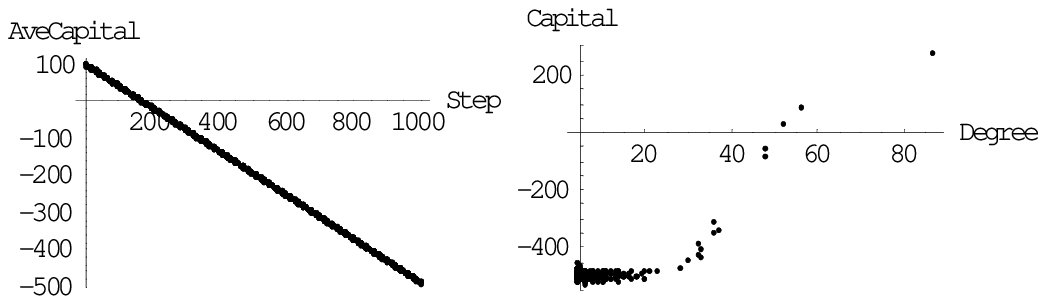} 
\end{center}
\caption{ Average capital $C(t)$ and capital distribution in every node in the game B at $R=10$ and $(P_L, P_R)=(0.2,0.6)$. }
\vspace{4mm}
\begin{center}
\includegraphics[scale=0.9,clip]{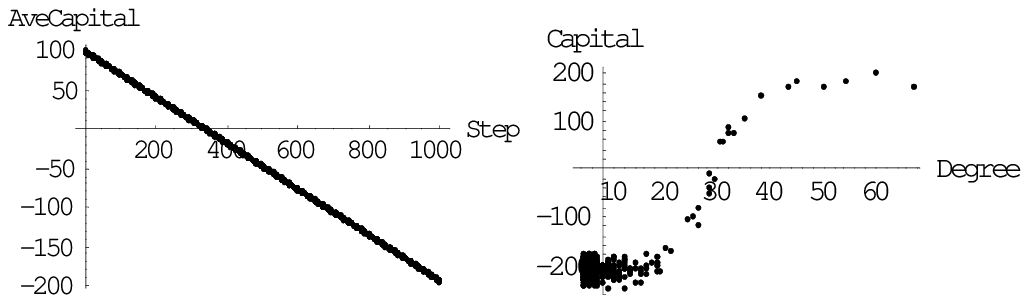} 
\end{center}
\caption{Average capital $C(t)$ and capital distribution in every node in the game A+B at $R=10$ and $(P_L, P_R)=(0.2,0.6)$.}
\end{figure}

\subsection{Numerical Analyses }
\hspace{5mm} 
We first put to the test by a computer simulation at $N=400$ and $P=0.5$  in order to discern the outline of wide parameter space. 
A necessary condition for the paradox is 
\begin{equation}
(P_L-0.5)(P_W-0.5)<0. 
\end{equation}
Computer simulations are made  under this parameter region where each player on the network asynchronously plays with $C(0)=100$.    
From eq.(32), we partition the parameter region into tow parts as follows; 
  \begin{enumerate}
\item $R=10$，$P_L>0.5>P_W$\\
 In this case, a typical time series of an average capital over all players and the capital distribution of each player at $t=1000$ 
 are shown in Fig.3 for the game B and Fig.4 for the game A+B. 
 As is expected, players on the nodes with low degree have much capital but hub players is not so in both games.   
In general, the average capital $C_B(t)$ in the game B is lager than  $C_{A+B}(t)$ in the game A+B. 
So it would be difficult that the average capital in the game A+B  exceeded the one in the game B. 

\item $R=10$，$P_L<0.5<P_W$\\
In this case, a typical time series of an average capital over all players and the capital distribution of each player at $t=1000$ 
 are shown in Fig.5 for the game B and Fig.6 for the game A+B. 
 As is expected, a few players on the nodes with high degree have much capital but players on the nodes with low degree is not so in both games.   
Since $C_{A+B}(t)>C_B(t)$ also applies in this case generally, it seems that a paradox may occur in some adequate parameter set.   
However,  $C_{A+B}(t) \propto C_B(t)$ applies at a rough estimate universally in various parameter sets, and 
just when the time variation $\delta C_{A+B}(t)$ of the lower capital $ C_{A+B}(t)$ 
turns into positive number,  the time variation $\delta C_B(t)$ of the larger capital $C_B(t)$ also turns into positive number. 
Thus we speculate the paradox does not occur in this case, too.  
\end{enumerate}

Furthermore, the variation of $P_L$ and $P_W$ within each parameter region is only relevant to the variation of the absolute values of $C_{A+B}(t)$ 
but does not occasion any qualitative changes such as a reversal in capital. 

We can speculate that decreasing $R$ makes losers increase in the case 1, but makes losers decrease 
in the case 2 from the data of the capital distribution in every node of Fig.3-6. 
Fig.7 and Fig.8 show $C(t) $ at $R=5$ in the game B and the game A+B, respectively. 
These figures verify that the speculation is right.  
These circumstantial evidence shows that the variation of $R$ contributes to the absolute value of the capital, and does not bring about any qualitative change 
such as the occurrence of paradox.  
We will make some discussions to support these in the next section. 

 \begin{figure}[t]
\begin{center}
\includegraphics[scale=0.9,clip]{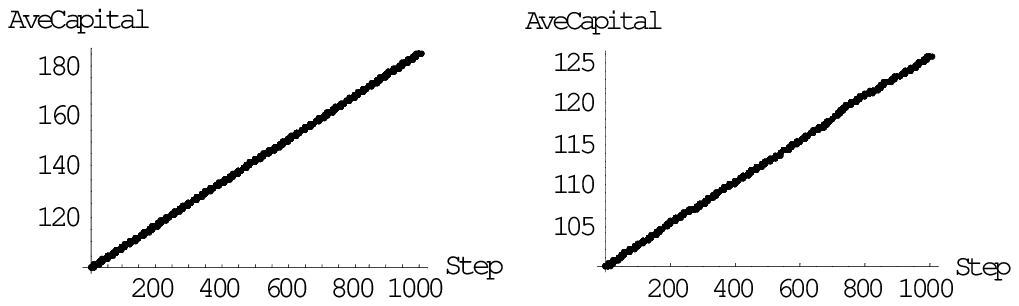} 
\end{center}
\caption{The average capital $C(t)$ at $R=5$ and $(P_L, P_R)=(0.6,0.2)$ in the game B and the game A+B.  }
\vspace{4mm}
\begin{center}
\includegraphics[scale=0.9,clip]{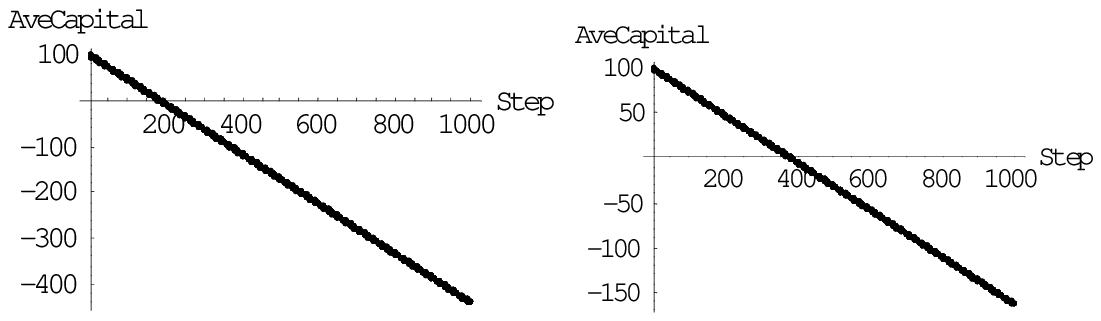} 
\end{center}
\caption{The average capital $C(t)$ at $R=5$ and $(P_L, P_R)=(0.2,0.6)$ in the game B and the game A+B.  }
\end{figure}

\subsection{Theoretical studies II}
\hspace{5mm}
In this case, it may be effective that the theoretical analysis made 
via DTMC(discrete-time Markov chains) such as given in \cite{Harm3},\cite{Miha1},\cite{Miha2}. 
 Every node, however,  has a different degree in scale free networks and so the size of the transfer matrix can not be fixed. 
 Therefore the analysis based on DTMC is difficult for scale free networks.   
 We make the analysis in this case based on the mean field approximation such as a recursion relation studied in the subsection 2.2. 
 
 According to \cite{Amen}, when a target paler plays different games depending on the number of winners adjacent to the target player, 
the discriminant of the node with degree $k$ is given by 
\begin{equation}
 D_{(k,p_i)}= \frac{ \displaystyle\prod_{i=0}^{k} (1-p_{i,B}) }{ \displaystyle\prod_{i=0}^{k} p_{i,B} },   
\end{equation}
where $p_{i,B}$ is the winning probability of $i+1$ one of $k+1$ games in the gameＢ. 
There are too parameters to analyze the game and consider a simpler case. 
As is explained in the subsection 3.1, we introduce a cutoff $R$.   
Then the discriminant for the node with degree $k$ for the game B in Parrondo's game on scale free networks based on analogical inference of eq.(23)
is 
\begin{equation}
D_{(k,P_{W,B} ,P_{L,B}  )}= \frac{ (1-P_{W,B})^{k-R} (1-P_{L,B} )^R }{ P_{W,B}^{k-R} P_{L,B}^{R} },   
\end{equation}
where $P_{1,B}=P_{W,B}$ and $P_{2,B} =P_{L,B} $ for $i=1,2$ ( $P_W$ and  $P_L$ that have already been used in this article are the short forms for them).
 
We need only to  replace the subscript $B$ with $A+B$  for the game A+B where  
the relations of winning probability between both games are given by eq.(15) and eq.(16).  

Let's consider the condition that paradox occurs according to the previous analysis.  
Notice that when the degree of  nodes  is smaller than the cutoff $R$, the players on the nodes necessarily play 
the game L in the game B. 
Then the discriminant is given by 
\begin{equation}
D_{(k,P_{L,B} )}= \frac{ (1-P_{L,B} )^k }{ P_{L,B}^{k} }.   
\end{equation}

Players on nodes with high degree can play either the game W or the game L corresponding to the number of winners in neighboring players.
Taking account of $k_{min} =4$ in Barabashi model ($\alpha=3$) adopted in this article, we find the condition for paradox;  
\begin{eqnarray}
\Bigl( \sum_{K=4}^R  \; \frac{D_{(K,P_{L,B} )}}{K^{\alpha}} &+&  \sum_{K=R+1}^k  \; \frac{ D_{(K,P_{W,B} ,P_{L,B}  )} }{K^{\alpha}}  \Bigl) 
\times   \frac{(\alpha -1)}{ k_{min}^{1-\alpha} - k_{max}^{1-\alpha} }>1, \\
 \Bigl( \sum_{K=4}^R  \;  \frac{D_{(K,P_{L,A+B} )}}{ K^{\alpha}  }&+&  \sum_{K=R+1}^k  \;  \frac{ D_{(K,P_{W,A+B} ,P_{L,A+B}  )}}{ K^{\alpha} } \Bigl) 
\times   \frac{(\alpha -1)}{ k_{min}^{1-\alpha} - k_{max}^{1-\alpha} } <1. 
\end{eqnarray}
In the continuous approximation, it becomes 
\begin{eqnarray}
\Bigl( \int_{4}^R  \; \frac{D_{(K,P_{L,B} )}}{ K^{\alpha} }dK&+&  \int_{R+1}^k   \frac{D_{(K,P_{W,B} ,P_{L,B}  )}}{ K^{\alpha} }dK \Bigl) 
\times   \frac{(\alpha -1)}{ k_{min}^{1-\alpha} - k_{max}^{1-\alpha} }<1, \\
 \Bigl( \int_{4}^R  \; \frac{D_{(K,P_{L,A+B} )}}{  K^{\alpha}  }dK&+&  \int_{R+1}^k   \frac{D_{(K,P_{W,A+B} ,P_{L,A+B}  )}}{ K^{\alpha} } dK\Bigl) 
\times   \frac{(\alpha -1)}{ k_{min}^{1-\alpha} - k_{max}^{1-\alpha} } >1. 
\end{eqnarray}
We will consider these conditions by making numerical calculations in the next section.

 \begin{figure}[t]
\begin{center}
\includegraphics[scale=0.9,clip]{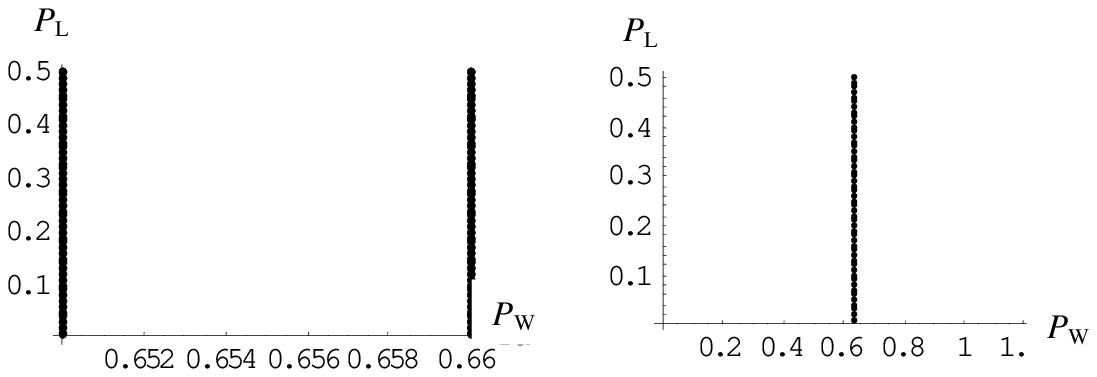} 
\end{center}
\caption{($P_W$，$P_L$) that satisfies the condition eq.(36) and (37) at $R=6$ for $k_{max}=25$ and $k_{max}=65$.  }
\vspace{4mm}
\begin{center}
\includegraphics[scale=0.9,clip]{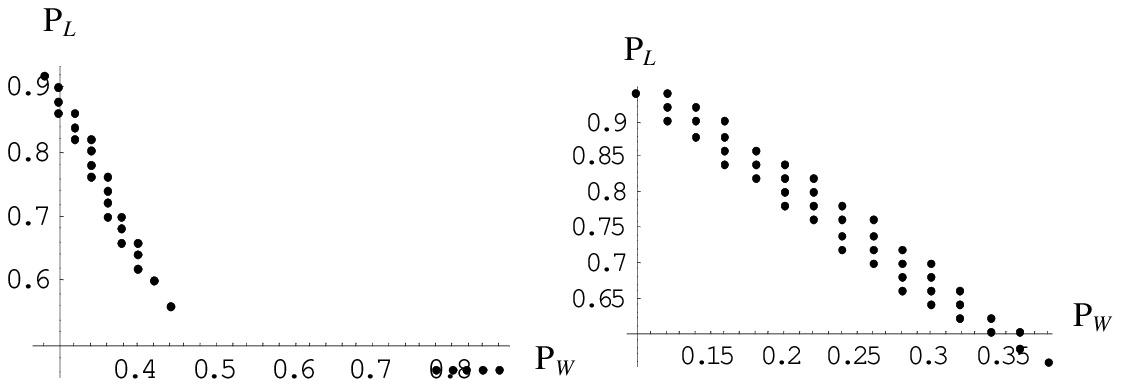} 
\end{center}
\caption{ ($P_W$，$P_L$) that satisfies the condition eq.(38) and (39) at $R=4$ and $R=9$ for $k_{max}=25$.    }
\begin{center}
\includegraphics[scale=0.9,clip]{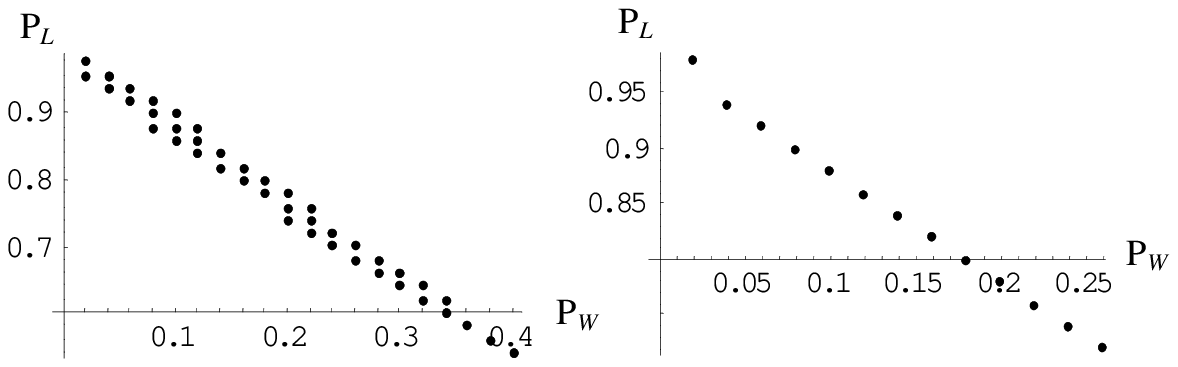} 
\end{center}
\caption{  ($P_W$，$P_L$) that satisfies the condition eq.(38) and (39) at $R=30$ for $k_{max}=65$ and at $R=73$ for $k_{max}=150$.  }
\end{figure} 

\subsection{Simulations }
\hspace{5mm}
The data points that satisfy  the conditions eq.(36) and (37) for discrete version in the parameter space $P_W$-$P_L$  are shown in Fig.9,  
where $P=0.5$ and $k_{max}= 25 (\mbox{ so }R<25)$ in the left hand side of Fig.9 and 
$k_{max}= 65 (\mbox{so} R<65)$ in the right hand side of Fig.9 are taken, respectively. 
The data points are estimated in 0.01 segments for both $P_L$ and $P_W$. 
The left hand side of Fig.9 corresponds to $N\sim400$ from eq. (31). 
We can confirm that these data points decrease as $k_{max}$ becomes larger by comparing the left and the right in fig.9
Furthermore since we can actually confirm that these data points disappear as the network size grows still larger, 
it is thought that they appear as the finite size effect.     
By way of precaution, trying actually making simulations at $N\sim400$ corresponding to $k_{max}=25$, we found that paradox does not occur in any parameters in the left hand side of Fig.9.     

Fig. 10 shows the data points that satisfy the conditions eq.(38) and eq. (39) in the continuous approximation.  
The left figure  is what is evaluated for small $R$ and the right figure is what is evaluated for relatively large.   
As result of simulations, we also found that paradox does not occur  on these data points at both $R$. 
Again we confirmed that the data points in the left hand side of Fig.10, which appear at small $R$, disappear as $k_{max}$ becomes larger. 
As for large $R$, data points appear in some straight lines shown in the right hand side of Fig.10. 
These points appear only at still larger $R$ as shown in Fig.11, as $k_{max}$ becomes still  larger. 
So it is conjectured that the data points in the straight lines disappear  at the limit of $N \rightarrow \infty$, which is a sort of the finite size effect. 
More simulation experiments, however, should be made to justify this statement. 
We need an electric  computer of rather high performance  to simulate Parrondo's game systematically at networks with $N>5000$ corresponding to $k_{max}\gtrsim 60$. 
It is a future problem whether paradox occurs at large $R$ in large scale networks such being the case.

\section{Sumamry}
\hspace{5mm}
In this article, we explored whether paradox occurs or not in Parrondo's game on scale free networks which are more ubiquitous in real worlds than regular networks.   
It is too complicate to analyze the game in the general fashion, especially giving theoretical considerations.  
So we consider only the case with the same number of parameters as the original Parrondo's game based on modulo $M=3$ in the capital.   
In our article, the parameter corresponding to $M$ in the original Parrondo's game  is the cutoff $R$.    
When the number of winners adjacent to a target player is less than $R$, 
the player plays the game L with the winning probability $P_{L}$ in the game B of Parrondo's game. 
Otherwise the player plays the game W with the winning probability $P_{W}$. 

First of all, we accumulated circumstantial evidence that paradox does not occur by some computer simulations.
Furthermore we practically prove that Parrondo's paradox does not occur in this naive case 
from theoretical point with numerical experiments of view in the final analysis. 
It, however, remains to be studied whether paradox actually does not occur in large scale networks. 

We never showed that Parrondo's paradox does not occur in scale free networks, generally. 
The networks that  we studied is only Barabashi model with the scaling exponent three and 
we only study excessively naive setting in game B.  
We only focus  on the numbers of winners, and not on degree or the number of losers
(notice that considering both the number of losers and winners turns  out to consider degree and the number of winners as well). 
If making efficient use of  these information, there would be sufficient possibilities that paradox occurs in various types of scale free networks.


\small

\end{document}